\documentclass[useAMS,usenatbib]{mn2e}

\usepackage{psfig}
\usepackage{amsbsy}
\usepackage{aas_macros}

\newcommand{\msun}{\ensuremath{ {\rm M}_{\odot} }}

\newcommand\etal{et al.\ }

\newcommand{\captionfonts}{\footnotesize\bf}
\makeatletter  
\long\def\@makecaption#1#2{%
  \vskip\abovecaptionskip
  \sbox\@tempboxa{{\captionfonts #1: #2}}%
  \ifdim \wd\@tempboxa >\hsize
    {\captionfonts #1: #2\par}
  \else
    \hbox to\hsize{\hfil\box\@tempboxa\hfil}%
  \fi
  \vskip\belowcaptionskip}
\makeatother   

\def\simlt{\lower.5ex\hbox{$\; \buildrel < \over \sim \;$}}
\def\simgt{\lower.5ex\hbox{$\; \buildrel > \over \sim \;$}}

\newcommand{\be}{\begin{equation}}
\newcommand{\ee}{\end{equation}}
\newcommand{\ba}{\begin{eqnarray}}
\newcommand{\ea}{\end{eqnarray}}

\newcommand{\llb}{\mbox{\boldmath $\ell$}}
\newcommand{\lgl}{\langle}
\newcommand{\rgl}{\rangle}

\title[Simulating Non-Linear Mode-Coupling in Parameter Estimation ]
      {Simulating the Effect of Non-Linear Mode-Coupling in Cosmological Parameter Estimation  }
      \author[Kiessling     \etal]    {A.     Kiessling\thanks{E-mail:
          aak@roe.ac.uk}, A.   N.  Taylor, A.   F.  Heavens\\ Scottish
        Universities Physics Alliance (SUPA), Institute for Astronomy,
        University   of   Edinburgh,   Royal  Observatory,   Blackford
        Hill,\\ Edinburgh, U.K.}

\begin{document}

\date{In preparation}

\pagerange{\pageref{firstpage}--\pageref{lastpage}} \pubyear{0000}

\maketitle

\label{firstpage}

\begin{abstract}
Fisher Information  Matrix methods are  commonly used in  cosmology to
estimate  the accuracy  that cosmological  parameters can  be measured
with   a   given  experiment,   and   to   optimise   the  design   of
experiments. However, the standard  approach usually assumes both data
and parameter estimates are Gaussian-distributed.  Further, for survey
forecasts  and optimisation  it is  usually assumed  the power-spectra
covariance  matrix   is  diagonal   in  Fourier-space.   But   in  the
low-redshift Universe, non-linear mode-coupling will tend to correlate
small-scale  power,  moving  information  from lower  to  higher-order
moments of  the field.  This  movement of information will  change the
predictions  of cosmological  parameter  accuracy.  In  this paper  we
quantify  this loss  of  information by  comparing na\"{i}ve  Gaussian
Fisher matrix forecasts with a Maximum Likelihood parameter estimation
analysis  of a  suite of  mock  weak lensing  catalogues derived  from
N-body  simulations, based  on the  SUNGLASS pipeline,  for a  2-D and
tomographic shear analysis  of a Euclid-like survey. In  both cases we
find that  the 68\% confidence  area of the  $\Omega_m-\sigma_8$ plane
increases by  a factor  5.  However, the  marginal errors  increase by
just 20  to 40\%.   We propose a  new method  to model the  effects of
nonlinear   shear-power  mode-coupling   in  the   Fisher   Matrix  by
approximating the shear-power  distribution as a multivariate Gaussian
with a covariance matrix derived from the mock weak lensing survey. We
find that this approximation can reproduce the 68\% confidence regions
of  the full  Maximum Likelihood  analysis in  the $\Omega_m-\sigma_8$
plane  to high  accuracy for  both  2-D and  tomographic weak  lensing
surveys.  Finally, we perform a multi-parameter analysis of $\Omega_m,
~\sigma_8,  ~h, ~n_s,  ~w_0$ and  $w_a$  to compare  the Gaussian  and
non-linear mode-coupled Fisher matrix contours.  The 6-D volume of the
$1\sigma$  error contours  for  the non-linear  Fisher  analysis is  a
factor of  3 larger than for the  Gaussian case, and the  shape of the
68\%  confidence volume is  modified.  We  propose that  future Fisher
Matrix estimates  of cosmological parameter  accuracies should include
mode-coupling effects.

\end{abstract}

\begin{keywords}
Gravitational  Lensing: Weak  -- Cosmology:  large scale  structure of
Universe -- Methods: Numerical, Statistical
\end{keywords}

\section{Introduction}
Measuring cosmological  parameters and the  equation of state  of dark
energy  to high  accuracy is  the goal  of many  current  and upcoming
experiments       (e.g.       PS1\footnote{PS1      http://ps1sc.org},
DES\footnote{DES                    https://www.darkenergysurvey.org/},
Euclid\footnote{Euclid  http://sci.esa.int/euclid}, HALO\footnote{HALO
  Rhodes     et    al.,     in     preparation},    LSST\footnote{LSST
  http://www.lsst.org/}           and           WFIRST\footnote{WFIRST
  http://wfirst.gsfc.nasa.gov}).  Variables such as the size and depth
of the survey (amongst other  things) have a significant effect on the
ability   of   a   survey   to  constrain   cosmological   parameters.
Consequently,   significant  effort  must   be  spent   in  accurately
predicting what  these telescopes  will see, well  before construction
begins. This will  allow us to both to influence  the design phase and
to understand the  capabilities of the instrument once  the design has
been set.  In  order to make predictions for  these upcoming missions,
statistical tools must  be used to estimate the  accuracy they will be
able to achieve.  The current  standard for prediction uses the Fisher
matrix methodology  \citep{tth97,f35}.  Traditionally, Fisher matrices
have  been generated  with  data covariance  matrices  that assume  an
underlying     Gaussian    matter    and     radiation    distribution
\citep{tth97,k97},  which  is  accurate  for CMB  estimates  when  the
Universe  was  still  linear.    However,  this  is  not  an  accurate
representation of the low-redshift  Universe at smaller scales.  Error
bars  generated using these  Gaussian assumptions  may be  biased when
compared  with those  generated  using methods  that  account for  the
non-Gaussian, non-linear nature of the Universe.

In  this paper  we compare  the $1\sigma$,  two-parameter  $\Omega_m -
\sigma_8$ error  estimates from  the full maximum  likelihood analyses
with  a Gaussian  Fisher analysis.   Using weak  gravitational lensing
power spectrum analysis of mock galaxy shear catalogues generated with
the {\small SUNGLASS} pipeline  \citep{kht+11}, we show the importance
of  using  accurate  non-linear  covariance matrices  when  estimating
errors  for  future   experiments.   Analytic  approximations  of  the
correlation  function covariance matrix  under these  assumptions have
been shown to underestimate the  errors on cosmic variance by a factor
of  up to  $\sim30$, which  makes breaking  the $\Omega_m  - \sigma_8$
degeneracy  more difficult  \citep{svh+07}.  Simulations  are  able to
provide  accurate  covariance  matrices   because  they  do  not  make
assumptions  about  the underlying  Gaussianity  of  the Universe  and
consequently,   the   resulting   covariance  matrices   include   the
off-diagonal components.

Fisher matrix analyses are  attractive because of their relative speed
and  minimal  computational requirements  when  compared with  maximum
likelihood  estimates.   However,  this  is  offset  by  the  loss  of
information  due to  Gaussian  assumptions in  the  generation of  the
Fisher matrix.  To compensate  for some of these Gaussian assumptions,
we propose  using a non-Gaussian weak lensing  shear covariance matrix
generated  from  simulations  to  generate  the  Fisher  matrix.   The
resulting  Fisher   matrix  still  assumes   a  multivariate  Gaussian
parameter  estimate  distribution  and  power  spectrum  distribution.
However,  it now  contains  the non-linear  information  found in  the
off-diagonal components of the covariance matrix from the simulations,
giving us a  `non-linear' Fisher matrix.  In this  paper, we will show
the  effect of  this simple  modification  to the  calculation of  the
Fisher matrix on the error estimates and compare this with the maximum
likelihood    and   Gaussian    Fisher   error    contours    in   the
$\Omega_m-\sigma_8$ plane. We  also perform a multi-parameter analysis
of $\Omega_m,~  \sigma_8, ~ h, ~n_s,  ~w_0$ and $w_a$  and compare the
Gaussian and non-linear Fisher matrix error estimates.

It  is possible  to produce  data vectors  for covariance  matrices by
performing   a   simple  2-D   binning   of   the   galaxies  in   the
survey. However,  further information may  be gained by  splitting the
distribution  up in  to  redshift bins  and  performing a  tomographic
analysis \citep[e.g.][]{h99,h02b,jt03}. In  this work, we perform both
2-D  and 3-bin tomographic  analyses in  order to  generate covariance
matrices that  are used to calculate the  maximum likelihood estimates
and the Gaussian and non-linear Fisher matrices.

The outline  of this paper is as  follows.  Section \ref{sec:sunglass}
will  detail  how the  simulations  and  mock  galaxy catalogues  were
generated. Section  \ref{sec:fisher} will introduce  the Fisher matrix
formalism   and    the   maximum   likelihood    formalism.    Section
\ref{sec:results}  shows the results  of the  analyses on  the maximum
likelihood estimates  and the Gaussian and  non-linear Fisher matrices
from the 2-D analyses in Section \ref{sec:unbinned} and the tomographic
analyses in Section  \ref{sec:tomo}.  Section \ref{sec:multi} compares
the   multi-parameter  Gaussian   and  non-linear   Fisher  estimates.
Finally,  a summary  of  the  findings will  be  presented in  Section
\ref{sec:disc}.

\section{Details of the simulations}
\label{sec:sunglass}

\begin{table}
  \begin{center}
    \begin{tabular}{cccccc}
      \hline\\  

      $N$ & Area (sq deg) & $z_{\rm max}$ & $n_g$/sq arcmin
      & $z_{\rm med}$ \\

      \hline\\ 

      100 & 100 & 1.5 & 15 & 0.82 \\ 

      \hline
    \end{tabular}
  \end{center}

  \caption{Table of  parameters for  the mock galaxy  shear catalogues
    used in this paper.  $N$  is the number of independent lightcones,
    $z_{\rm  max}$  is  the  maximum  redshift in  the  lightcone  and
    $n_{g}$/sq arcmin is the number of `galaxies' per square arcmin in
    the  catalogue and  $z_{\rm med}$  is the  median redshift  of the
    catalogue. The  suite of  lightcones is used  together to  form a
    survey with an effective area of 10,000 sq. deg.}
  \label{tab:lightconetable}

\end{table}

The suite of weak lensing  simulations used in this work was generated
using   the   {\small  SUNGLASS}   pipeline   \citep[for  a   detailed
  introduction, see][]{kht+11}.   We have 100  independent simulations
generated  with  the  cosmological  N-body  simulations  code  {\small
  GADGET2}  \citep{sp05}.   The  simulations  were made  with  a  flat
concordance  $\Lambda$CDM cosmology, consistent  with the  WMAP 7-year
results  \citep{jbd+11}:  $\Omega_m   =  0.272$,  $\Omega_{\Lambda}  =
0.728$,  $\Omega_b =  0.045$, $\sigma_8  = 0.809$,  $n_s =  0.963$ and
$h=0.71$ in  units of 100  km s$^{-1}$ Mpc$^{-1}$.  There  are $512^3$
particles in a  box of $512h^{-1}$~Mpc which leads  to a particle mass
of $7.5 \times 10^{10}\msun$.  The simulations were all started from a
redshift  of $z=60$  and  allowed to  evolve  to the  present with  26
snapshots being stored in redshifts $0.0 \le z \le 1.5$.

Lightcones  were   generated  through  the   simulation  snapshots  to
determine the  average convergence in  an angular pixel using  the `no
radial binning' method introduced in \cite{kht+11}:
\begin{equation}
\kappa_p(r_s) = \sum_k \frac{K(r_k,r_s)}{\Delta\Omega_p \bar{n}
(r_k)r_k^2} - \int_0^{r_s}\!dr~ K(r,r_s),
\end{equation}
where $\bar{n}$ is the number  density of particles in the simulation,
$\Delta\Omega_p  = \Delta\theta_x\Delta\theta_y$  is  the pixel  area,
$r_s$ is the comoving radial  distance of the lensing source plane and
$r_k$ is  the comoving radial distance  of each particle,  $k$, in the
lightcone. $K(r,r_s)$ is the scaled lensing kernel:
\begin{equation}
K(r,r_s) = \frac{3 H_0^2 \Omega_m}{2c^2} \frac{(r_s - r)r}{r_sa(r)}.
\end{equation}
The  convergences were  calculated  using $2048^2$  azimuthal bins  on
source redshift planes  that were separated by $z =  0.1$ to create 15
planes from  $0.0 < z <  1.5$. Once the  convergences were calculated,
shear  values were  determined in  Fourier space,  where the  shear is
given by $\gamma = \gamma_1 + i\gamma_2$.

The shear and convergence values  in these source redshift planes were
interpolated back  onto the individual particles in  the lightcones to
generate mock  galaxy shear  catalogues. The B-modes  (the unphysical,
imaginary component  of the convergence)  in the mock  catalogues were
calculated directly from the shear. These catalogues were then sampled
to    reproduce    a    standard    galaxy    redshift    distribution
\citep[e.g.][]{ebt+91},
\begin{equation}
n(z) \propto  z^\alpha \exp \left[  - \left(\frac{z}{z_0}\right)^\eta
  \right],
\end{equation} 
where $z$ is  the redshift, $z_0$, $\alpha$ and  $\eta$ set the depth,
low-redshift  slope  and  high-redshift  cut-off for  a  given  galaxy
survey.  We  take $\alpha=\eta=2$ and  $z_0=0.78$ in this  work, which
gives a median redshift $z_{med} = 0.82$.  It is assumed that galaxies
trace the dark matter distribution perfectly and the final mock galaxy
shear catalogues  contain 15 galaxies per square  arcminute.  There is
no ellipticity noise in the  catalogues, however there is a shot-noise
contribution related to the discrete  sampling of the particles in the
mock catalogues.   Table \ref{tab:lightconetable} summarises  the mock
galaxy catalogues used in this work.

\section{Methodology}
\label{sec:fisher}

The  shear,  $\gamma$,  convergence,  $\kappa$, and  B-mode,  $\beta$,
fields are related to each other in Fourier-space on a flat-sky by
\begin{equation}
\kappa(\llb) +  i\beta(\llb) = e^{2 i  \varphi_\ell} [\gamma_1(\llb) +
  i\gamma_2(\llb) ],
\end{equation}  
where $\varphi_\ell$  is the angle between the  angular wave-vector and
an axis on the sky.  For  each of the mock galaxy shear catalogues the
shear, convergence, and B-mode auto- and cross-power spectra have been
estimated.  The tomographic weak lensing shear, convergence and B-mode
power cross-spectra, for two  different source redshifts $z$ and $z'$,
are given by
\begin{eqnarray}
C_\ell^{\gamma\gamma} (z,z') &=& \langle \gamma_1(\boldsymbol{\ell},z)
\gamma_1(\llb,z')  \rangle   +  \langle  \gamma_2(\boldsymbol{\ell},z)
\gamma_2(\llb,z') \rangle,\\
C_\ell^{\kappa\kappa}(z,z')                 &=&                \langle
\kappa(\boldsymbol{\ell},z)\kappa(\llb,z') \rangle,\\
C_\ell^{\beta\beta}(z,z')   &=&   \langle   \beta(\boldsymbol{\ell},z)
\beta(\llb,z') \rangle.
\end{eqnarray}
The power spectra are  related to each other by $C_\ell^{\gamma\gamma}
(z,z')= C_\ell^{\kappa\kappa}  (z,z')+ C_\ell^{\beta\beta}(z,z')$. The
auto-spectra  are  calculated when  $z'=z$.  In  practice  we bin  the
sources into redshift slices.  For  a survey that has $N_z$ slices the
expectation-value of the tomographic  shear power spectrum in redshift
bins labeled $i$ and $j$, is given by
\begin{equation}
C^{\gamma\gamma}_{ij}(\ell)       =      \frac{9H_0^4\Omega_m^2}{4c^4}
\int^{r_{\rm max}}_0  \!  \frac{dr}{a^2(r) }~  P\left(\frac{\ell}{r} ,
r\right) g_{i}(r) g_{j}(r),
\end{equation}  
\citep{k92,js09}, where $r$ is the comoving distance, $r_{\rm max}$ is
the  maximum comoving  distance  and $P(\ell/r,r)$  is  the 3D  matter
density power spectrum. The lensing efficiency function is
\begin{equation}
g_{i}(r)  =   \int_r^{r_{\rm  max}}  \!dr~  p_{i}  (r')   \left(  1  -
\frac{r}{r'} \right),
\end{equation}
where $p_{i}$ is the normalised probability distribution of `galaxies'
in the bin.

When  determining  the angle  averaged  shear  power  spectrum in  the
simulations, we  must take into consideration the  conventions used in
the  Fourier  transform  software  FFTW\footnote{The  Fastest  Fourier
  Transform  in the West  http://www.fftw.org}. Thus,  the discretised
tomographic shear power spectrum becomes
\begin{equation}
\frac{\ell (\ell +  1) \widehat{C}_{ij}^{\gamma\gamma}(\ell)} {2\pi} =
\!\!\!  \sum_{\ell \,  in \, shell} \!\!\!\!  \frac{\gamma_1(\llb,z_i)
  \gamma_1(\llb,z_j)   +   \gamma_2(\llb,z_i)\gamma_2(\llb,z_j)}{n_b^2
  \Delta \ln \ell},
\end{equation}
where $\widehat{C}_\ell^{\gamma\gamma}$ is  the estimated power, $n_b$
is the  total number of  bins in the  Fourier transform, $z_i$  is the
$i^{th}$ redshift slice,  and $\Delta \ln \ell$ is  the thickness of a
shell  in log  $\ell$-space.  The  modes  in this  power spectrum  are
corrected  for mode  discreteness errors  by scaling  by  the expected
number  of modes.   To compactify  the  notation we  shall denote  the
tomographic shear power by $C^{\gamma\gamma}_X(\ell)$, where $X=(i,j)$
is a pair  of redshift slices. For our 2-D analysis  there is only one
bin so $i=j$.  Using the information from these power spectra, we have
the the information required to make cosmological parameter estimates.

\subsection{Shear power covariance matrix generation}

The shear  cross-spectra covariance matrix,  $M^{XX'}_{\ell \ell'}$ is
defined by
\begin{equation}     
M^{XX'}_{\ell\ell'}  =  \lgl  \Delta  C^{\gamma\gamma}_X(\ell)  \Delta
C^{\gamma\gamma}_{X'} (\ell') \rgl,
\end{equation}
where  $\Delta  C^{\gamma\gamma}_X(\ell)=C^{\gamma\gamma}_X(\ell)-\lgl
C^{\gamma\gamma}_X(\ell)\rgl$  and  angled-brackets  denotes  ensemble
averaging.  The covariance matrix is an important element of parameter
estimation, containing information on the strength of the correlations
between variates,  in this case  the shear power spectrum  modes.  The
accuracy  of  this  matrix  improves  by  increasing  the  numbers  of
realisations  included in  the calculation  of the  matrix.   For this
work,  100 independent realisations  of 2-D  power spectra  provides an
accuracy  of $\Delta  C_\ell \sim  10\%$ \citep[as  shown in  Figure 7
  of][]{kht+11}.

\subsubsection{Gaussian covariance matrix}

If  the shear  field is  isotropic and  we assume  the shear  field is
Gaussian we can write down an equation for the tomographic shear power
covariance matrix  and make a correction  for the fraction  of the sky
covered by the survey:
\begin{eqnarray} 
M^{ij,kl}_{\ell\ell'}     \!\!\!     \!\!     &     =&    \!\!\!\!\!\!
\frac{\delta^K_{\ell\ell'}}    {(2\ell+1)    f_{\mathrm{sky}}}   \Big(
     [C_{ik}^{\gamma\gamma}       (\ell)+N_i(\ell)      \delta^K_{ik}]
     [C_{jl}^{\gamma\gamma}(\ell)+N_j(\ell)  \delta^K_{jl}]  \nonumber
     \\
  &   &   \,\,\,   +   \,\,\,   [C_{il}^{\gamma\gamma}(\ell)+N_i(\ell)
  \delta^K_{il}][C_{jk}^{\gamma\gamma}(\ell)+N_j(\ell)   \delta^K_{jk}]
\Big) ,
\label{eq:theoryCoV}
\end{eqnarray}
where $f_{\mathrm{sky}}$  is the  fraction of the  sky covered  by the
survey and $N_\ell$ is a  shot-noise term due to intrinsic ellipticity
in the  shear field.  For the  purposes of this work  although we have
discrete galaxy density,  we set the intrinsic shear  variance to zero
so that  $N_\ell=0$. In principle our  results can be  combined with a
Gaussian noise covariance to model different weak lensing surveys.

\subsubsection{Simulation covariance matrix estimation}
 The covariance matrix of the tomographic shear power spectrum for the
 suite of mock galaxy shear catalogues is estimated by
\begin{equation}
M^{XX'}_{\ell\ell'}     =     \frac{1}     {N-1}     \sum_N     \Delta
C_{X}^{\gamma\gamma}(\ell) \Delta C_{X'}^{\gamma\gamma} (\ell'),
\label{eq:CoV}
\end{equation}
where  $N$ is  the  number  of mock  catalogue  realisations and  here
$\Delta  C_{X}^{\gamma\gamma} (\ell)  = \widehat{C}_{X}^{\gamma\gamma}
(\ell) -  \lgl \widehat{C}_{X}^{\gamma\gamma}(\ell) \rgl$  where $\lgl
\widehat{C}_{X}^{\gamma\gamma}(\ell) \rgl$ is  the ensemble average of
the tomographic shear power spectrum across all realisations.

We  calculate  inverse  covariance   matrix  using  a  singular  value
decomposition (SVD) on  the covariance matrix \citep{ptv+92}. However,
the resulting inverse is biased due  to noise. To correct for this, we
multiply the inverse by a factor \citep{hss07}:
\begin{equation}
[\widehat{M}^{XX'}_{\ell\ell'}]^{-1} = \frac{N_S -  N_p - 2} {N_S - 1}
[M^{XX'}_{\ell\ell'}]^{-1},
\end{equation}
where $N_S$ is  the number of realisations, $N_p$  is the total number
of     bins     in     all     of    the     power     spectra     and
$[\widehat{M}^{XX'}_{\ell\ell'}]^{-1}$  is an  unbiased  estimation of
the inverse covariance matrix.

\subsection{Fisher matrix generation}

The Fisher matrix is given by

\begin{equation}
F_{ij}  =  \frac{1}{2} \sum_{\ell  \ell'}  \sum_{XX'} \frac  {\partial
  C^{\gamma\gamma}_X(\ell)}       {\partial      \theta_i}      \left[
  \widehat{M}^{XX'}_{\ell\ell'}\right]^{-1}    ~    \frac    {\partial
  C^{\gamma\gamma}_{X'}(\ell')} {\partial \theta_j} ,
\label{eq:Fisher}
\end{equation}
where $i$ and $j$  label cosmological parameters (e.g.  $\Omega_m$ and
$\sigma_8$)  and  the partial  derivatives  are  the  gradient of  the
expectation value of the shear  power spectrum in parameter space.  In
order  to determine  that gradients  of the  power spectra,  we  use a
five-point function  \citep[][equation 25.3.6]{as68} and  the ensemble
average  power  spectra.  The  theoretical  prediction  for the  power
spectrum  was generated  using a  code provided  by  Benjamin Joachimi
\citep[as tested in][]{kht+11,js09}.  This code uses the \cite{spj+03}
non-linear  power spectrum estimate,  the \cite{eh98}  matter transfer
function  and  a numerically  calculated  linear  growth factor.   All
further  theoretical power  spectrum  predictions in  this paper  were
performed using this same code.

The Fisher  matrix provides  parameter estimates through  two methods.
The first by  calculating the area of the  Fisher matrix error ellipse
that  encloses  a  two-parameter  68\%  confidence limit  in  the  two
parameter  plane. The  inverse of  this  area is  proportional to  the
Figure-of-Merit     that     is     often    quoted     in     studies
\citep[e.g.][]{abc+06,w08,aab+09,sl10}.     The    second    is    the
single-parameter marginal errors, which are given by
\begin{equation}
\Delta \theta_i = \left[F^{-1}\right]_{ii}^{1/2}.
\end{equation}

\subsection{Maximum likelihood parameter estimation}
To perform  a maximum likelihood analysis  on the suite  of mock shear
catalogues, we use a  Gaussian likelihood estimator.  Despite the fact
that the simulations are non-Gaussian, using a Gaussian likelihood has
been shown to produce accurate results \citep{kht+11}.  The likelihood
is given by
\begin{equation}
L(\hat{C}_\ell^{\gamma\gamma}|\sigma_8,\Omega_m)       =      \frac{1}
{(2\pi)^{N/2}   ({\rm   det}  \widehat{M}^{XX'}_{\ell   \ell'})^{1/2}}
\exp\left[- \frac{\chi^2} {2} \right],
\label{eq:Like}
\end{equation}
where $N$  is the number  of independent mock  catalogue realisations,
and
\begin{equation}
\chi^2  =  \sum_{\ell  \ell'} \sum_{XX'}  \Delta  C_{X}^{\gamma\gamma}
(\ell)      \left[\widehat{M}^{XX'}_{\ell\ell'}\right]^{-1}     \Delta
C_{X'}^{\gamma\gamma}(\ell'),
\end{equation}
where        $\Delta        C_{X}^{\gamma\gamma}       (\ell)        =
\widehat{C}_X^{\gamma\gamma}(\ell)   -   \langle  C_{X}^{\gamma\gamma}
(\ell)  \rgl$,  and  $\langle  C^{\gamma\gamma}_\ell \rangle$  is  the
expected  angular power spectrum  given by  equation (8).   While this
likelihood  analysis  is  computationally  expensive,  it  takes  into
account the full non-Gaussian and non-linear nature of the simulations
and  should provide the  most accurate  error estimates.   The maximum
likelihood is related to the Fisher matrix by
\begin{equation}
F_{ij}  =  \left \langle  -  \frac{\partial^2\ln L}{\partial  \theta_i
  \partial \theta_j}\right\rangle.
\end{equation}

\section{Results}
\label{sec:results}
\subsection{2-D two-parameter analysis}
\label{sec:unbinned}
\begin{figure}
  \psfig{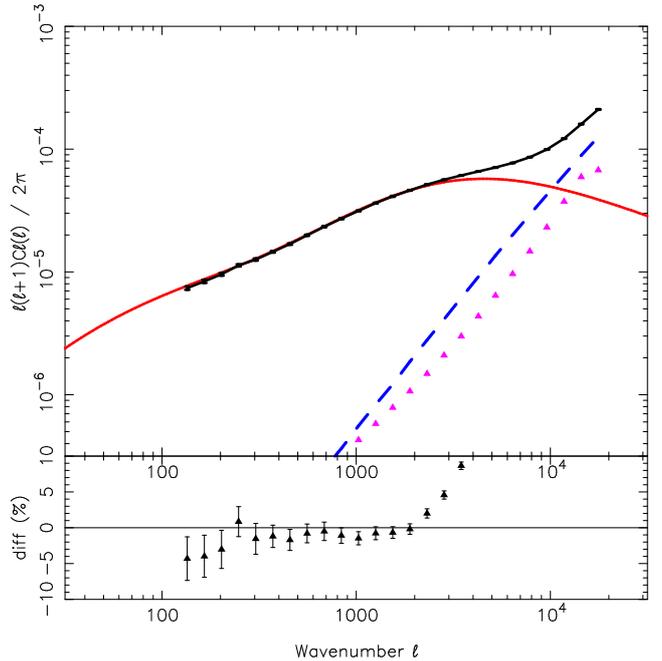}
  \caption{Mean 2-D shear power spectrum for the suite of unbinned mock
    galaxy shear  catalogues.The smooth (red) line  is the theoretical
    prediction for the power spectrum, the (black) line shows the mean
    power spectrum for the suite of mock catalogues with errors on the
    mean shown.  The  diagonal (dark blue) dashed line  shows the shot
    noise  prediction and  the (magenta)  triangles show  the measured
    B-modes. The  bottom panel shows the the  percentage difference of
    the data from the expected power spectrum with errors.}
  \label{fig:nzPS}
\end{figure}

The  first step  to making  parameter estimates  from the  mock galaxy
shear catalogues  is to determine  the 2-D shear angular  power spectra
for  each  realisation. Figure  \ref{fig:nzPS}  shows  the mean  power
spectra  from  the  suite  with   the  (red)  long  line  showing  the
theoretical  prediction, the  (black) line  showing the  mean measured
power spectrum from  the suite of catalogues with  errors on the mean.
The (magenta) triangles show the measured B-modes and the dashed (dark
blue)  line  shows  the  shot-noise  arising  from  discrete  particle
sampling  in  the  mock   catalogues.   The  bottom  panel  shows  the
percentage difference  of the data  from the expected  power spectrum.
The simulations recover the expected power spectrum within 5\% between
$150 < \ell < 2000$.

The  shot noise  prediction  for these  catalogues  was determined  by
filling a  suite of simulation volumes with  randomly placed particles
to  mimic Poisson  noise. The  {\small SUNGLASS}  pipeline was  run on
these noise simulations and the  power spectrum of the mock catalogues
is the dashed line in the figure. We know that the shot-noise in these
catalogues is not purely Poissonian because the simulations start from
a glass pre-initial distribution \citep{w94}. However, the noise tends
toward Poissonian as the simulations  evolve. We are not able to model
this  noise  accurately because  it  evolves  with  structures in  the
simulation  \citep{bge95}. Thus,  this  Poisson estimate  should be  a
reasonable approximation for the noise in these catalogues.

A data-vector  was constructed from  the shear band-power  spectra with
wavenumbers  from $150 <  \ell <  1500$. From  this, we  estimated the
covariance  matrix in equation  (\ref{eq:CoV}). Similarly,  a Gaussian
covariance matrix was generated using equation (\ref{eq:theoryCoV}).

To  test the  effect of  the  non-Gaussian nature  of the  simulations
covariance matrix,  we looked at  the diagonal components of  both the
simulation and Gaussian field covariance matrices and determined their
$\Delta C_{\ell}$  values.  For  the case of  the simulations  this is
simply the square  root of the diagonal components.   For the Gaussian
field this is
\begin{equation}
\Delta C_\ell  = \frac{\sqrt{2}C_\ell}{\sqrt{\ell(2\ell+1) f_{\rm  sky} \Delta
    \ln \ell}}.
\end{equation}

\begin{figure}
  \centerline{\psfig{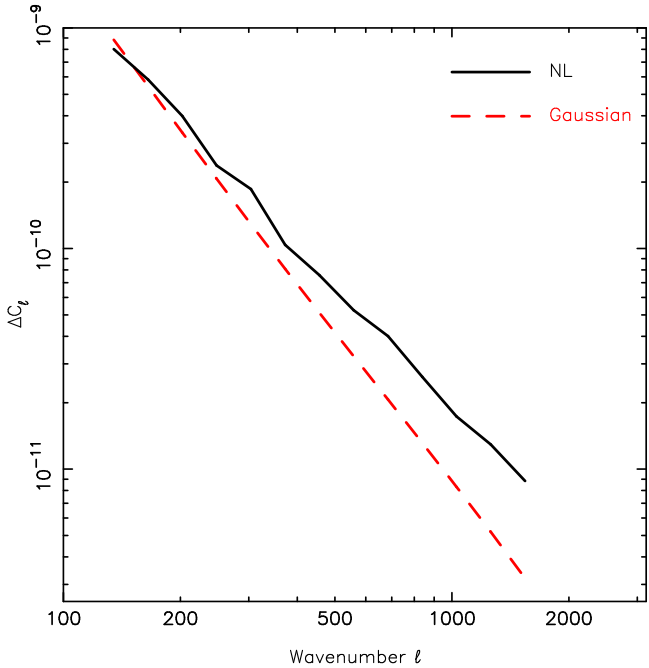}}
  \caption{Diagonal  components  of  the  covariance  matrix  for  the
    simulations (black line) and a  Gaussian field with the same power
    spectrum (red dashed line).}
  \label{fig:CoVError}
\end{figure}

Figure  \ref{fig:CoVError}  shows  the  diagonals  of  the  covariance
matrices as a function of wavenumber $\ell$.  The (red) dashed line is
from  the  Gaussian  error  and  the (black)  line  is  from  the
simulations.   At low wavenumbers,  which is  still in  the reasonably
linear regime, the two errors  agree reasonably well.  However, in the
higher  wavenumber, non-linear  regime,  there is  a  factor of  three
difference  between  the  errors,  suggesting  that  the  non-Gaussian
contribution to the covariance is significant.

Fisher  matrices in the  $\Omega_m -  \sigma_8$ plane  were calculated
using equation (\ref{eq:Fisher}) for  both the Gaussian and simulation
covariance matrices.  The Fisher matrices were multiplied  by 100, the
total  number of  independent  mock catalogues,  to  provide an  error
estimate  for a survey  of 10,000  square degrees.   Additionally, the
maximum  likelihood  was calculated  using  the simulation  covariance
matrix and the likelihoods from  the 100 mock catalogues were combined
\citep[as shown  in ][]{kht+11}. For  all of these calculations  it is
assumed that all other cosmological parameters are known.

\begin{figure}
  \centerline{\psfig{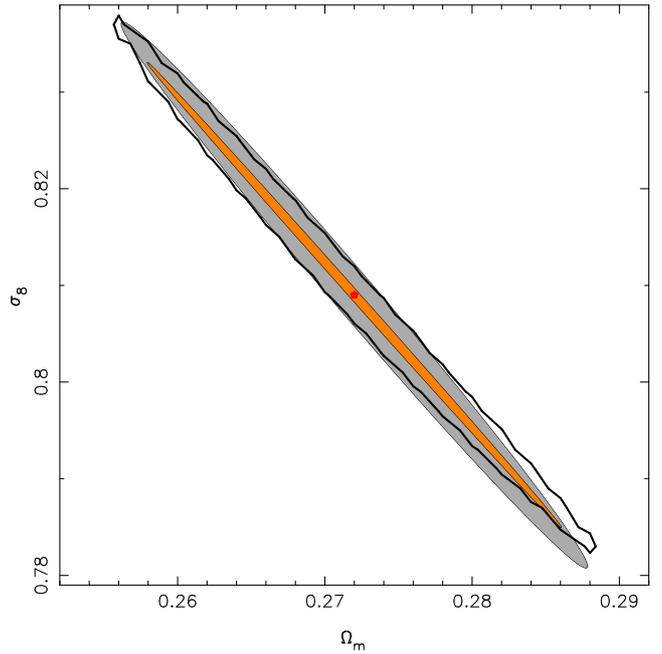}}
  \caption{Comparison of the  $1\sigma$ two-parameter contours for the
    $\Omega_m-\sigma_8$ parameters for  Gaussian Fisher contour (inner
    orange shaded ellipse) with  the non-linear Fisher contour (middle
    grey  shaded ellipse)  and  the full  maximum likelihood  analysis
    (thick black line  contour). The (red) point in  the middle of the
    ellipses represents the fiducial $\Omega_m-\sigma_8$ parameters in
    this calculation.}
  \label{fig:LikeFish}
\end{figure}

In Figure \ref{fig:LikeFish}  the two-parameter $1\sigma$ contours for
the $\Omega_m - \sigma_8$ plane  are shown.  The inner (orange shaded)
ellipse shows  the Gaussian Fisher  estimate, the outer  (grey shaded)
ellipse  shows the non-linear  Fisher estimate  and the  thick (black)
line  contour is the  combined maximum  likelihood estimate.   The red
point in the middle  represents the fiducial $\Omega_m$ and $\sigma_8$
parameters. The  size and shape  of the simulation Fisher  ellipse and
the  maximum likelihood  contour are  very similar,  showing  that the
non-linear  Fisher calculation  is  a good  method  for making  future
cosmological parameter estimates,  provided the off-diagonal terms are
included. The  area of  the non-linear Fisher  estimate is  5.1 times
larger than the area of the  Gaussian Fisher estimate which is a clear
indication that  the off-diagonal terms  in the covariance  matrix are
essential for accurate estimation of the parameter errors.

In  addition  to the  area  of the  contours  we  also determined  the
marginal  errors on  the  parameters from  the  Fisher matrices.   The
non-linear  $\Omega_m$ marginal error  is 1.23  times larger  than the
Gaussian marginal  errors and $\sigma_8$  is 1.17 times  larger.  From
Figure \ref{fig:CoVError},  we would expect the marginal  errors to be
much  larger.   However,  the  off-diagonal terms  in  the  simulation
covariance matrix act  to reduce the marginal errors  but increase the
total area. This is demonstrated  by setting the off-diagonal terms in
the  simulation  covariance matrix  are  to  zero.  In this  case  the
non-linear Fisher ellipse becomes far narrower but the marginal errors
increase significantly.

\begin{figure}
  \centerline{\psfig{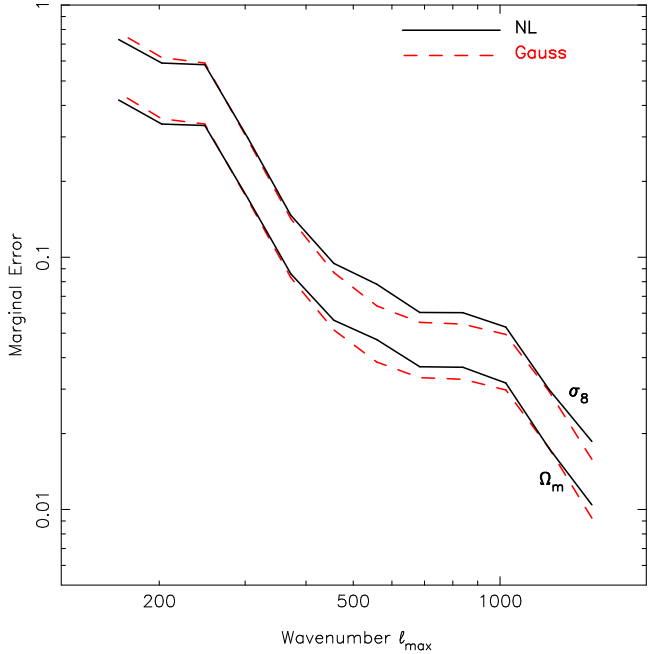}}
  \caption{Marginal errors of $\Omega_m$  and $\sigma_8$ as a function
    of maximum wavenumber in  the Fisher calculation. The dashed lines
    are the  marginal errors  on $\Omega_m$ as  a function  of maximum
    wavenumber  $\ell_{\rm  max}$ and  the  continuous  lines are  for
    $\sigma_8$.}
  \label{fig:Marginal}
\end{figure}

Figure \ref{fig:Marginal} shows how  the marginal errors change in the
non-linear  Fisher   matrix  as  a  function   of  maximum  wavenumber
$\ell_{\rm  max}$.  The marginal  errors for  both the  non-linear and
Gaussian Fisher  matrices are remarkably similar across  all values of
$\ell_{\rm max}$  with the largest  gain of information  occurring for
both $\Omega_m$  and $\sigma_8$ between  $250 < \ell_{\rm max}  < 500$
and again  at $\ell_{\rm max}  > 1000$. This  similarity is due  to an
effect that  the off-diagonal components of  the simulation covariance
matrix are  having on  the error ellipse,  as discussed  earlier. This
figure shows that if the marginal  error is the value of interest, the
Gaussian Fisher matrix appears to  provide a result that is comparable
with the error  obtained with the non-linear Fisher  error estimate at
wavenumbers between $150 < \ell_{\rm  max} < 1500$.  However, when the
errors are marginalised over two  parameters, the area of the contours
shows that the errors are being underestimated by a factor of five.

\subsection{Tomographic shear power}
\label{sec:tomo}

It  is possible to  perform the  same analysis  that was  performed in
Section \ref{sec:unbinned}  on tomographically-binned mock catalogues.
Tomography introduces  extra information  into the analysis.   In this
analysis  the suite  of mock  galaxy shear  catalogues was  split into
three redshift bins; bin  1: $0.0 \le z \le 0.5$, bin  2: $0.5 < z \le
1.0$ and bin 3: $1.0 < z \le 1.5$.

\begin{figure*}
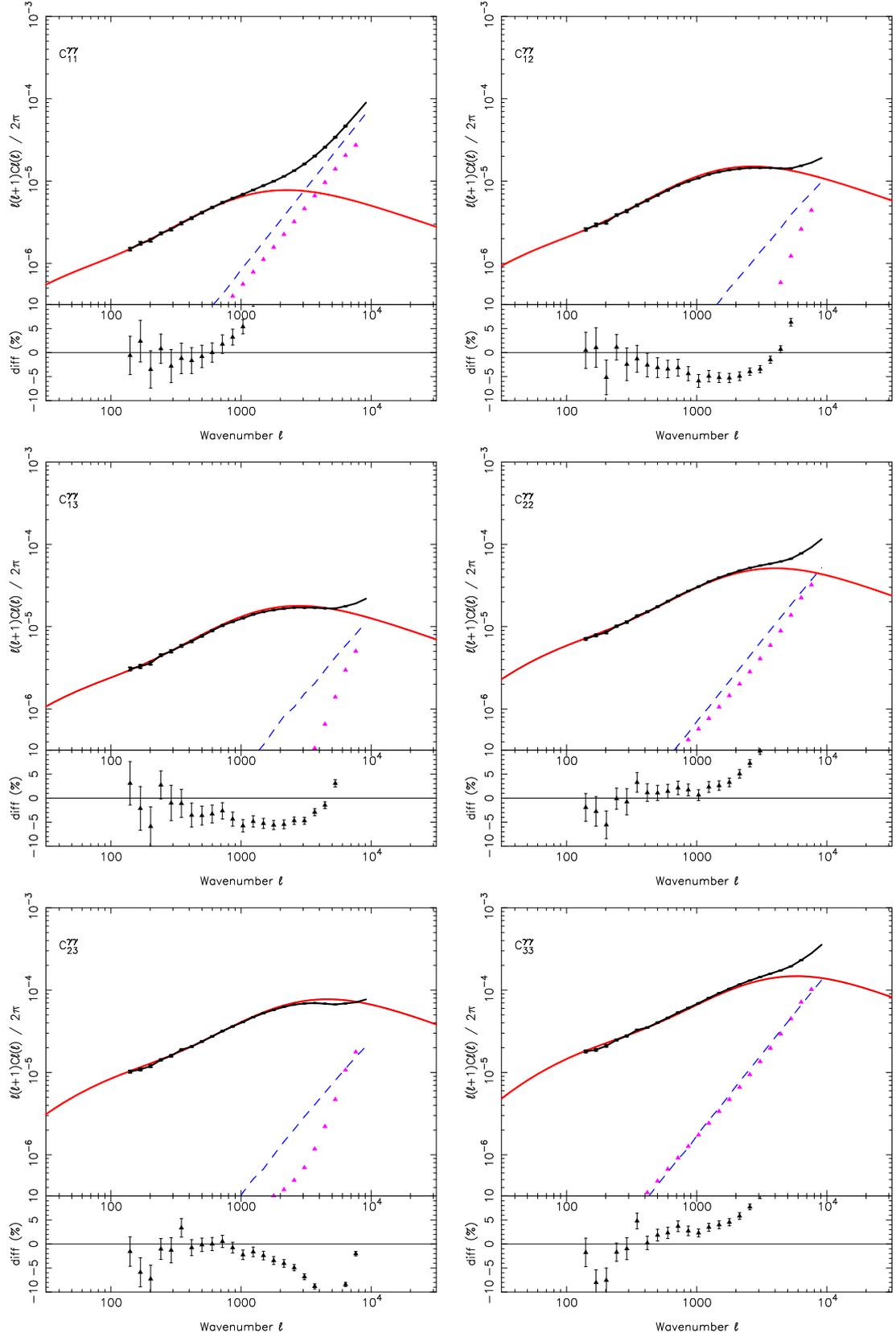

\begin{minipage}{148mm}
  \begin{tabular}{cc}
    \psfig{file=Tomo_PS_1.ps,angle=-90,width=0.475\linewidth,clip=} &
    \psfig{file=Tomo_PS_2.ps,angle=-90,width=0.475\linewidth,clip=}\\

    \psfig{file=Tomo_PS_3.ps,angle=-90,width=0.475\linewidth,clip=} &
    \psfig{file=Tomo_PS_4.ps,angle=-90,width=0.475\linewidth,clip=}\\

    \psfig{file=Tomo_PS_5.ps,angle=-90,width=0.475\linewidth,clip=} &
    \psfig{file=Tomo_PS_6.ps,angle=-90,width=0.475\linewidth,clip=}

  \end{tabular}
\end{minipage}
\caption{2-D shear power spectra for a 3-bin tomography analysis of the
  mock galaxy shear catalogues. The long (red) line is the theoretical
  prediction for  the shear  power spectrum, the  (black) line  is the
  mean power spectrum  for the 100 mock catalogues  with errors on the
  mean, the (dark blue) dashed line is the shot noise estimate and the
  (magenta) triangles are the measured B-modes. }
\label{fig:PSComp}
\end{figure*}

Figure \ref{fig:PSComp}  shows the  auto- and cross-power  spectra for
the 3-bin  tomographic analysis of  the mock galaxy  shear catalogues.
If we  first focus  on the auto-power  spectra ($C_{11},~  C_{22}$ and
$C_{33}$), we  can see that  the wavenumbers that we  model accurately
increase  with   redshift.   For  $C_{11}$,  we   find  agreement  for
wavenumbers from $150 < \ell  < 700$.  The power spectra from $C_{22}$
and $C_{33}$ are slightly higher  than the expected power spectrum but
still  within  3\%  and  are  reliable  up to  $\ell  =  1000$  before
shot-noise becomes dominant.

The cross-bins $C_{12}, ~C_{13}$ and  $C_{23}$ are damped which is due
to the shot  noise dominance in the lower  redshift bin. Consequently,
we only use the results of  these power spectra up to the $\ell$ range
accurately recovered  by the auto-power spectra of  the lower redshift
bin.

The  power spectra  in the  auto- and  cross-bins were  turned  into a
data-vector with  wavenumbers from  $150 < \ell  < 1800$  included.  A
data covariance  matrix was generated  and from this,  the correlation
coefficient matrix where the correlation coefficients are given by
\begin{equation}
r^{XX'}_{\ell\ell'}       =       \frac{\widehat{M}^{XX'}_{\ell\ell'}}
{\sqrt{\widehat{M}^{XX}_{\ell\ell} \widehat{M}^{X'X'}_{\ell'\ell'}}}.
\end{equation}
This matrix shows how (anti-)correlated each of  the $\ell$ modes
are. 

\begin{figure*}
\centering
  \psfig{file=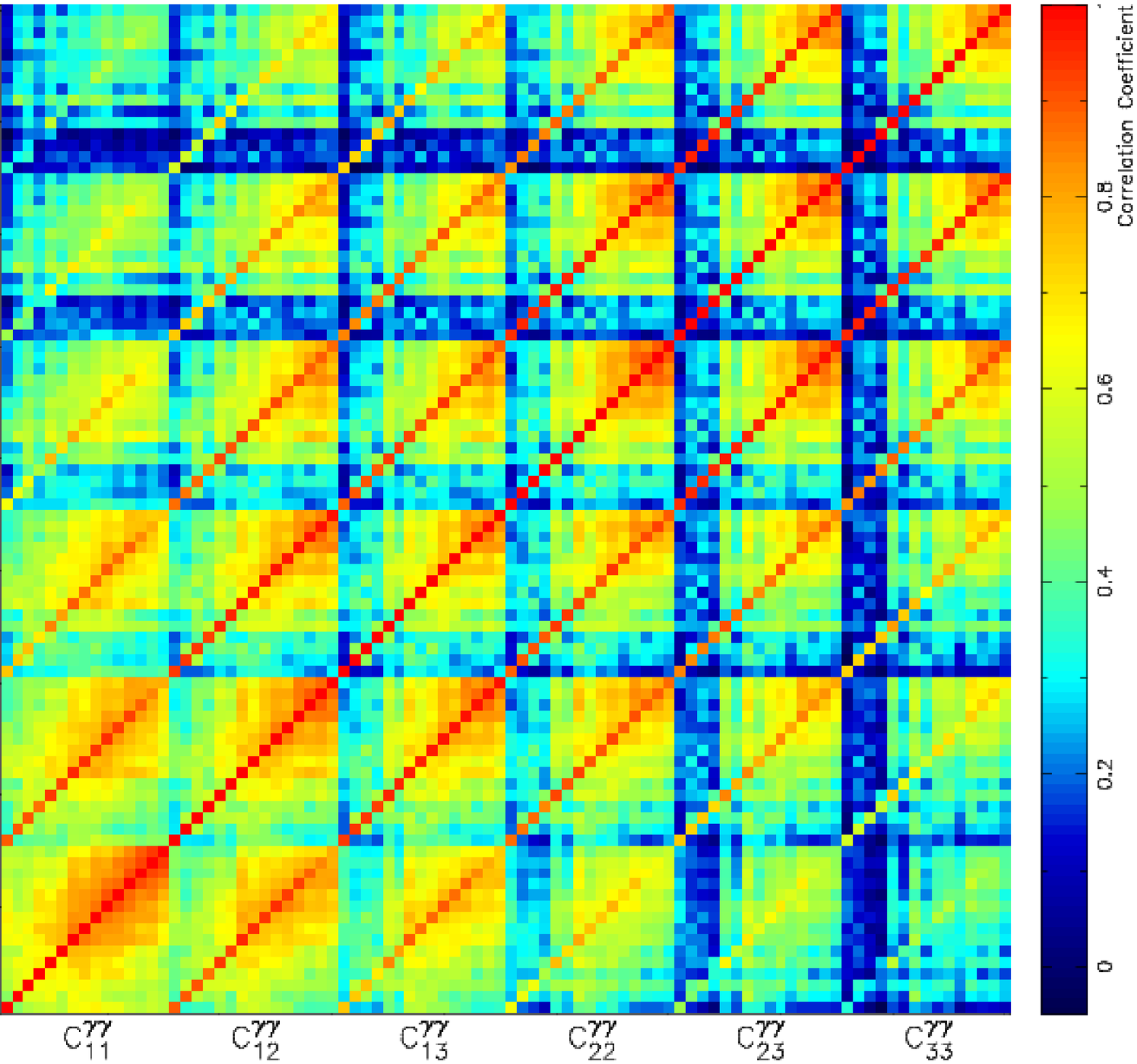,angle=-0,width=\linewidth,clip=}
  \caption{Correlation coefficient matrix  for 3 tomographic bins. The
    matrix is  arranged so that  each sub-square is a  tomographic bin
    pair.}
  \label{fig:CCM_Tomo}
\end{figure*}

Figure \ref{fig:CCM_Tomo} shows the correlation coefficient matrix for
the  tomographic  data  vector.   The  `block' nature  to  the  matrix
indicates each  tomographic bin  pair.  Each of  the blocks  along the
diagonal   represents  the  auto-correlation   between  each   of  the
tomographic power spectrum analyses.  The off-diagonal blocks show the
cross-correlations  between the  tomographic pair  power  spectra.  As
expected, the higher wavenumbers  in each block are highly correlated.
Additionally, the  $C_{11}$ auto-power spectrum  is highly correlated.
The $C_{22}$ auto-power spectrum  is significantly less correlated and
the $C_{33}$ auto-power spectrum bin has very low correlations between
the $\ell$ modes.   The indication of this is  that the lower redshift
tomographic  bins should  contribute fewer  power spectrum  bins  to a
covariance matrix being used for any kind of analysis.

With  the  information  gained   from  these  power  spectra  and  the
correlation coefficient matrix, we  are able to assemble a data-vector
that  contains  the  $C_\ell$  from wavenumbers  that  are  accurately
reproducing  the expected  power spectrum  and have  a  reasonably low
correlation.   Thus,  we  select  30  bins in  total  from  the  three
auto-power spectra,  with the highest wavenumber being  around $\ell =
1000$ and just 10 bins in total from the cross-power spectra, with the
highest wavenumber being around $\ell = 500$.  A covariance matrix was
generated using  this data-vector. A full  maximum likelihood analysis
was  performed with  this covariance  matrix and  a  non-linear Fisher
matrix was also calculated.   An equivalent Gaussian covariance matrix
was also generated and the Gaussian Fisher matrix determined.

\subsection{Two-parameter tomographic analysis}

\begin{figure}
  \centerline{\psfig{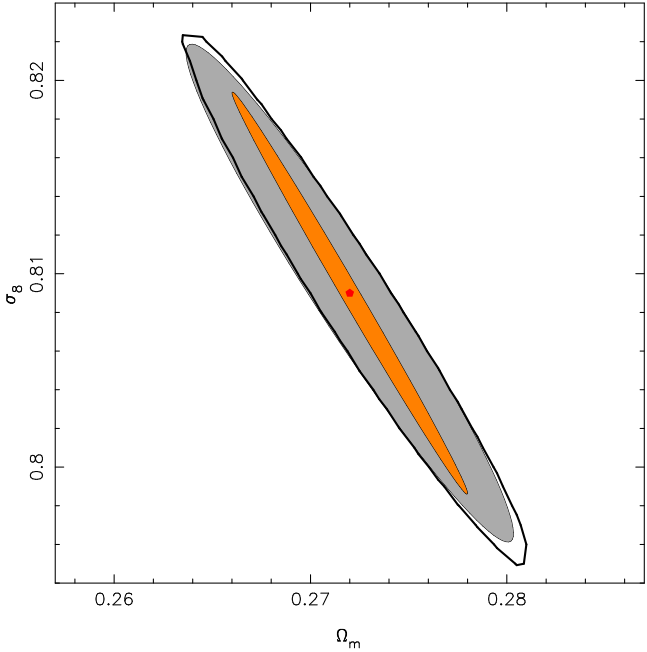}}
  \caption{Comparison of the $1\sigma$,  two-parameter contours for the
    $\Omega_m-\sigma_8$  parameters for  the  Gaussian Fisher  contour
    (inner  orange  shaded  ellipse),  the simulation  Fisher  contour
    (middle  grey  shaded ellipse)  and  the  full simulation  maximum
    likelihood analysis (thick black line contour). The (red) point in
    the   middle    of   the   contours    represents   the   fiducial
    $\Omega_m-\sigma_8$ parameters in this calculation.}
  \label{fig:LikeFish_Tomo}
\end{figure}

Figure  \ref{fig:LikeFish_Tomo}  shows  the  $1\sigma$,  two-parameter
contours  for the  Gaussian  Fisher (orange  shaded  ellipse) and  the
non-linear Fisher  (grey shaded ellipse).  The thick  black line shows
the simulation  maximum likelihood analysis.   As in the case  for the
2-D  analysis, the  non-linear Fisher  contour  is very  close to  the
maximum likelihood contour.  The area of the non-linear Fisher contour
is 5.1 times larger than the area of the Gaussian Fisher contour.  The
marginal error of the non-linear  Fisher estimate is 1.39 times larger
in $\Omega_m$ and 1.24 times  larger in $\sigma_8$.  These numbers are
very similar to those calculated in the 2-D analysis.

The area  of the contours and the  size of the marginal  errors in the
tomographic Fisher  and likelihood estimates is smaller  than those in
the  unbinned analyses,  showing  that the  tomographic analysis  does
indeed contain more information.

\subsection{Multi-parameter Fisher analysis}
\label{sec:multi}

\begin{table}
  \begin{center}
    \begin{tabular}{l|ccc}
      \multicolumn{4}{c}{\textbf{Tomographic Multi-Parameter Analysis}}\\
      \hline 
       & NL & Gauss & Gauss / NL \\

      \hline\\ 

      $\Delta \Omega_m$ & 0.009 & 0.010 & 1.11 \\ 
      $\Delta \sigma_8$ & 0.019 & 0.021 & 1.11 \\
      $\Delta h$        & 0.198 & 0.352 & 1.78 \\
      $\Delta n_s$      & 0.143 & 0.201 & 1.41 \\
      $\Delta w_0$      & 0.107 & 0.129 & 1.21 \\
      $\Delta w_a$      & 0.343 & 0.584 & 1.70 \\
      \hline
    \end{tabular}
  \end{center}

  \caption{Marginal  errors  of  a  3-bin  tomographic,  six-parameter
    Fisher matrix  analysis using both Gaussian  and non-linear Fisher
    matrices. For this configuration, the Gaussian marginal errors are
    always  larger than  the non-linear  Fisher errors.   However, the
    non-linear  Fisher volume is  3.7 times  larger than  the Gaussian
    volume.}
  \label{tab:marginal}

\end{table}

In the  previous section, we showed  that a Fisher  matrix analysis in
the $\Omega_m  - \sigma_8$ plane  accurately determines the  errors on
the parameters when  a non-Gaussian data covariance matrix  is used in
the calculation.  This  was shown by comparing the  68\% error contour
with a maximum  likelihood analysis of the simulation  suite.  In this
Section  we perform  a Fisher  matrix analysis  over  the cosmological
parameters: $\Omega_m, ~\sigma_8, ~h, ~n_s, ~w_0$ and $w_a$. Note that
$\Omega_\Lambda  = 1  -  \Omega_m$.  We  compare  the Gaussian  Fisher
analysis with the non-Gaussian (simulation) Fisher analysis.  Based on
the findings in the two-parameter  analysis, we assume that the Fisher
errors  calculated  with the  simulation  data  covariance matrix  are
consistent with a maximum  likelihood analysis of the simulations over
the same multi-parameter space.

\begin{figure*}
  \centerline{\psfig{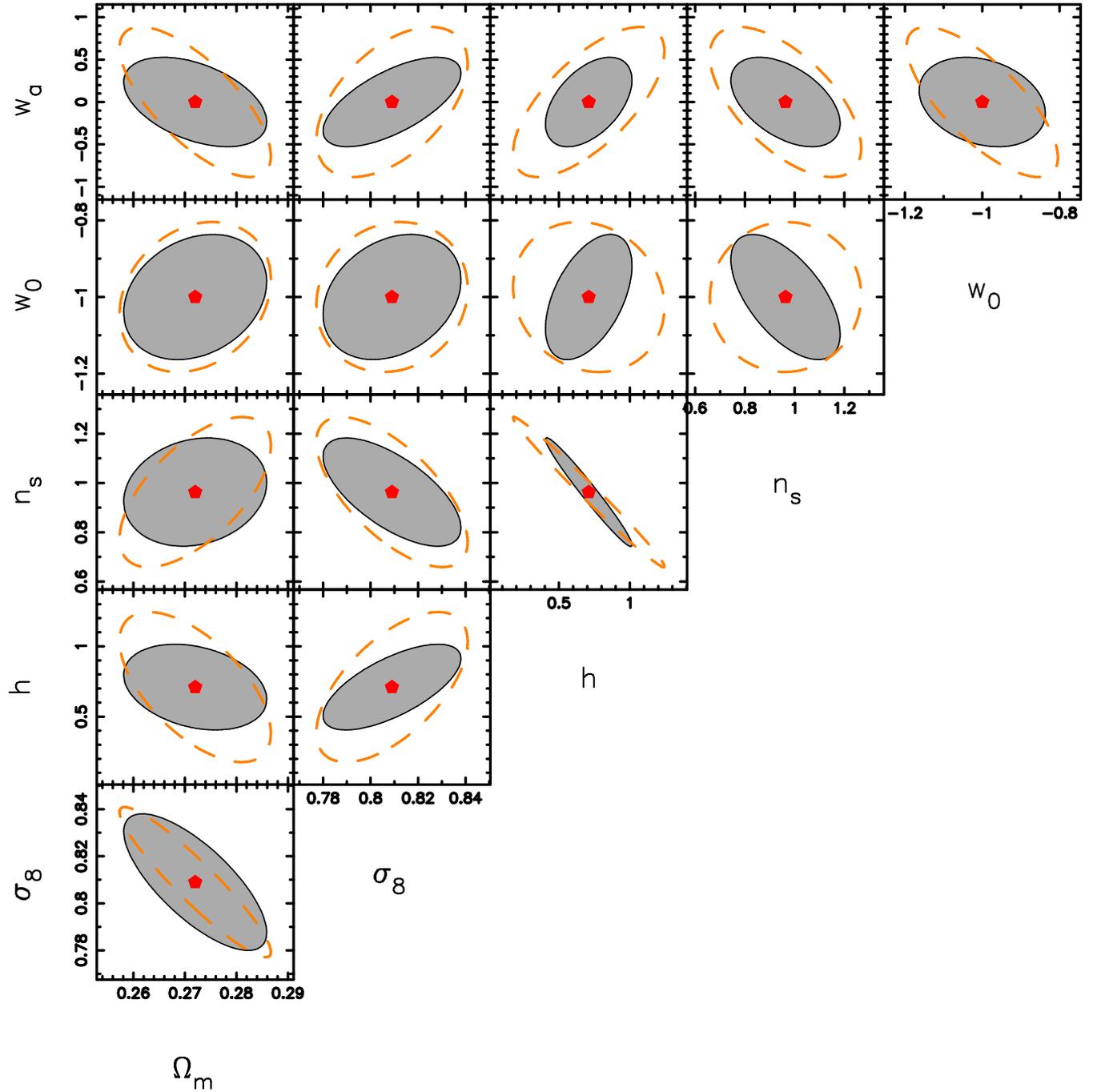}}
  \caption{Multiple  parameter  Fisher  analysis.  The  (grey)  shaded
    ellipse is  the non-linear Fisher contour and  the (orange) dashed
    ellipse is the Fisher contour  for the Gaussian Fisher matrix. The
    volume  of the  6-dimensional space  is larger  in  the simulation
    Fisher  matrix,  however some  of  the  projected contours  appear
    larger  for  the  Gaussian  Fisher  matrix.  This  could  be  very
    misleading.}
  \label{fig:Tomo_Multi}
\end{figure*}

For  a Euclid-like survey  (Table \ref{tab:lightconetable}),  it makes
little sense  to perform a  2-D analysis over multiple  parameters. We
found that the size of the  marginal errors was so large as to provide
no constraining information (e.g. $\Delta  w_0 = 11.3$ and $\Delta w_a
= 49.8$).  However, the  additional information provided by performing
a  tomographic  analysis  of  the  power spectra  yielded  far  better
constraints  and  illuminated  some  features  of  the  Fisher  matrix
analysis    that    were     not    immediately    obvious.     Figure
\ref{fig:Tomo_Multi}  shows  the  projected  two-parameter,  $1\sigma$
contours,  marginalised  over the  the  multiple  parameters from  the
Fisher  matrix analysis.   The (red)  point  at the  center shows  the
fiducial parameters  used in the analysis, the  (grey) shaded ellipses
show the  non-linear Fisher analysis and the  dashed (orange) ellipses
show  the  Gaussian  Fisher   analysis.   In  all  cases,  except  the
$\Omega_m-\sigma_8$ plane, the projected area of the Gaussian contours
is larger than the area  of the non-linear Fisher contours which gives
the  impression that  the  error estimates  from  the Gaussian  Fisher
analysis are more conservative the non-linear estimates.  The marginal
errors  of  the Gaussian  contours  are  larger  in all  cases  (Table
\ref{tab:marginal}).   However,  volume  of  the  six-parameter  space
generated,
\begin{equation}
V \propto \sqrt{\det(F^{-1})},
\end{equation}
shows that the  non-linear Fisher volume is 3.7  times larger than the
Gaussian volume, which is what  we would expect given the off-diagonal
terms  included  in  the   simulation  data  covariance  matrix.   The
implication  of this  is  that  although the  projected  areas of  the
Gaussian Fisher appear larger, the overall volume is smaller. This can
be explained  with an example  (in three dimensions for  simplicity of
explanation): Take a  spherical ball and a thin  plate with a slightly
larger radius than the  ball.  If we look at the ball  from any of the
three  axes, the  area  will appear  to  be the  same  circle in  each
projection.  If  we place the plate  at a $45^\circ$ angle  to each of
the three  axes and then look at  it in projection, it  will appear to
have a larger  area than the ball in all  projections, however we know
that the volume of the plate  is far smaller than the ball.  With this
knowledge in hand, it is easy to see how misleading the projections of
the Fisher matrices can be in these complex multiple-parameter spaces.

The  eigenvalues   and  eigenvectors  of  the   Fisher  matrices  were
determined  to  find  the  equation   for  the  plane  with  the  most
information in the full six-parameter space. For the Gaussian, this
is:
\begin{eqnarray}
X  &=&  0.865~\Omega_m  +  0.498~\sigma_8  +  0.049~n_s  +  0.030~h  -
\nonumber\\ && 0.035~w_0 - 0.006~w_a,
\end{eqnarray}
and for the simulations:
\begin{eqnarray}
Y  &=&  0.841~\Omega_m  +  0.536~\sigma_8  +  0.034~n_s  +  0.028~h  -
\nonumber \\ & & 0.056~w_0 - 0.015~w_a.
\end{eqnarray}
When  using the  fiducial parameters,  
\begin{eqnarray}
X & = & 0.741, \\
Y & = & 0.771,
\end{eqnarray}
and the errors  on the planes are $\Delta X =  8.6 \times 10^{-8}$ and
$\Delta  Y = 1.4  \times 10^{-6}$.   This shows  an order-of-magnitude
difference in the thickness of the 6-D error ellipses.

\section{Discussion and Conclusions}
\label{sec:disc}
This paper makes a comparison between Fisher matrices in the $\Omega_m
- \sigma_8$  plane and  six-parameter  Fisher matrices  ($\Omega_m,~
\sigma_8,  ~h,  ~n_s, ~w_0$  and  $w_a$),  generated using  covariance
matrices  from   a  Gaussian  random   field  and  from   full  N-body
simulations.    The   $1\sigma$   two-parameter  contours   from   the
two-parameter  $\Omega_m-\sigma_8$ Fisher  matrices are  also compared
with the contour from a  maximum likelihood analysis of the full suite
of simulations.

This  work  uses  the  {\small  SUNGLASS}  pipeline  to  generate  100
independent simulations of  512$h^{-1}$~Mpc with $512^3$ particles and
a   standard  $\Lambda$CDM  cosmology.    The  pipeline   turns  these
simulations into  100 independent mock galaxy shear  catalogues of 100
square degrees  and a  galaxy redshift distribution  with a  median of
$z_m  = 0.82$  and  15  `galaxies' per  square  arcminute. When  these
catalogues  are combined,  they provide  an effective  survey  area of
10,000 sq. deg.

We perform both 2-D and 3-bin tomographic angular shear power spectrum
analyses  on  each of  the  mock  catalogues  and generate  covariance
matrices from the resulting data-vectors.  We also generate a Gaussian
field   covariance  matrix   to  compare   with  the   more  realistic
non-Gaussian analyses.

Using these  covariance matrices, we generate Fisher  matrices for the
$\Omega_m - \sigma_8$ plane and show that the $1\sigma$, two-parameter
contour for the non-linear Fisher matrix has an area that is 5.1 times
larger than the theoretical prediction in both the 2-D and tomographic
analyses.    This  indicates  that   the  theoretical   prediction  is
significantly  under-predicting the  errors on  these  parameters (the
Figure-of-Merit is over-optimistic), even though their marginal errors
are similar.

To quantify  if the contours generated  using the Fisher  matrix are a
reasonable estimate of the true errors, we compared these outputs with
a combined  maximum likelihood analysis of the  full simulation suite.
The  resulting  contours in  both  the  2-D  and tomographic  analyses
closely  matched the  contours  generated with  the simulation  Fisher
matrix.   From this,  we can  conclude that  it is  sensible to  use a
Fisher  matrix analysis  for parameter  estimates, using  a covariance
matrix with all off-diagonal terms included.

Based on the  success of the non-linear Fisher  errors in matching the
maximum likelihood estimates in the $\Omega_m-\sigma_8$ plane, we also
perform  both   2-D  and   3-bin  tomographic  analyses   to  generate
multi-parameter  Fisher matrices  of $\Omega_m,  ~\sigma_8,  ~h, ~n_s,
~w_0$ and  $w_a$.  We  compare the non-linear  Fisher matrix  with the
Gaussian Fisher matrix under the assumption that the simulation Fisher
matrices  are providing  accurate  error estimates.   With the  survey
parameters used in this paper,  the 2-D analysis finds marginal errors
on the parameters that are  so large that they provide no constraining
power.

The   tomographic  analyses  find   reasonable  marginal   errors  and
demonstrate the value of  the additional information obtained from the
tomography.   However,  the marginal  errors  of  the Gaussian  Fisher
matrix are larger than the non-linear Fisher marginal errors for every
parameter  which  is a  counterintuitive  result.   We  show that  the
projected  contours from the  Gaussian Fisher  matrix are  also larger
than  the non-linear  Fisher matrix,  even  though the  volume of  the
non-linear Fisher is 3.7 times  larger than the Gaussian Fisher.  This
warns us  that the  projected Fisher contours  can be  misleading over
complex  multi-variate spaces  and  that larger  2-D  contours do  not
necessarily indicate a larger error volume.

\section*{Acknowledgments}
AK would  like to thank Benjamin  Joachimi and Tom  Kitching for their
very useful  discussions on this  work, the European DUEL  RTN project
MRTN-CT-2006-036133  and the University  of Edinburgh  for studentship
support.

\bibliographystyle{mn2e} 
\bibliography{thesis}

\end{document}